\documentclass[preprint,prb]{revtex4}
\usepackage{amsmath}
\usepackage{xcolor,color,soul}
\usepackage[utf8]{inputenc}
\usepackage{url}

\begin{document}

\title{Elementary Solution of Kepler Problem \\ (and a few other problems)}
\author{M.~Moriconi}
\email{mmoriconi@id.uff.br}
\affiliation{Universidade Federal Fluminense, Instituto de Física, Niterói, RJ, Brasil.}

\begin{abstract}
{\noindent We present a simple method to obtain the solution of a few orbital problems: the Kepler problem, the modified Kepler problem by the addition of an inverse square potential and linear force.}\\
{\bf Keywords:} classical mechanics, Kepler problem, celestial mechanics
\end{abstract}

\maketitle

\section{Introduction}

There is a handful of non trivial problems in classical mechanics one can solve exactly, such as the one-dimensional harmonic oscillator (1dHO) and the Kepler problem\cite{Symon, French, Morin, Chen, Davis, Vogt, Milnor, Simha, Unruh}. The latter one, that is, to find the orbit of a point mass under the action of a inverse square force, is particularly important for its historical role as well as applications in celestial mechanics.

Besides their important applications in the real world, these problems also serve as theoretical playground where one can apply different methods an techniques, with varying degrees of sophistication, as can be easily learned from analytical mechanics books \cite{Fasano, Lemos}.

In this note we present yet another solution by means of a very simple trick, using minimal technical background, avoiding cumbersome manipulations or involved mathematical calculations, making the approach suitable for beginners.

Initially we solve the 1dHO as a way to illustrate the method. Then we move to the Kepler problem, followed by the study of the Kepler problem modified by the addition of a $1/r^2$ potential and, at the end we discuss the orbit of a point mass under a quadratic potential which is, surprisingly, a bit more involved than the Kepler problem.

\section{Warmup Problem: $1d$ Harmonic Oscillator}

Consider a $1d$ harmonic oscillator, a point mass $m$ attached to a spring of constant $k$. This is a conservative system and we know its total mechanical energy, $E$, is conserved and is given by
\begin{equation}
E = \frac{m}{2} {\dot x}^2 + \frac{k}{2} x^2 . \label{1dHO}
\end{equation}
We can rewrite equation (\ref{1dHO}) as
\begin{equation} 
\left( \sqrt{\frac{m}{2E}} \dot x\right)^2 + \left(\sqrt{\frac{k}{2E}} x \right)^2 = 1.
\end{equation}
Whenever the sum of the squares of two real quantities adds up to $1$ we can parametrize them by trigonometric functions, namely, we can write each term 
 as
\begin{eqnarray}
    \sqrt{\frac{m}{2E}} \dot x &=& \sin \phi ,\label{sinEq} \\ \sqrt{\frac{k}{2E}} x &=& \cos \phi  .\label{cosEq}
\end{eqnarray}
where $\phi$ is a function of time, to be determined. Taking the time derivative of equation (\ref{cosEq}) and using equation (\ref{sinEq}), we obtain
\begin{equation}
    \sqrt{\frac{k}{2E}} \dot x = -\dot \phi \sin \phi = -\dot \phi \sqrt{\frac{m}{2E}} \dot x,
\end{equation}
which implies
\begin{equation}
\dot \phi =   - \sqrt{\frac{k}{m}} = -\omega ,
\end{equation}
where we introduced
$\omega = \sqrt{k/m}$. Therefore $\phi = - \omega t + \phi_0$ and the solutions for $x$ is
\begin{equation}
x =  \sqrt{\frac{2E}{k}} \cos(\omega t - \phi_0) .
\end{equation}
This is the complete solution of the $1d$ harmonic oscillator. Notice that the amplitude is already expressed in terms of the total energy of the system. The only adjustable parameter is the phase $\phi_0$, which can always be set to zero, by demanding our clock starts at the moment the particle is most distant from the origin.

\section{The Kepler Problem} \label{KeplerProblem}

Explaining the motion of celestial bodies, in particular the orbits of comets and planets in the solar system, is one of the greatest triumphs of Newtonian mechanics. This is a problem that deserves to be studied carefully and from as many different angles as possible. 

Let us treat the Kepler problem, that is, the motion of a point particle of mass $m$ orbiting a planet of mass $M$ under the central potential $V(r) = - GMm/r$.

Initially, we note that the angular momentum $\vec L = \vec r \times \vec p$ is conserved for central forces, for
\begin{equation}
    \frac{d}{dt} \vec L = \frac{d}{dt} \vec r \times \vec p + \vec r \times \frac{d}{dt} \vec p = \vec v \times \vec p + \vec r \times \vec F = 0
\end{equation}
since the velocity $\vec v$ is parallel to the momentum $\vec p$ and the position vector $\vec r$ is parallel to the central force $\vec F = F(r) \hat r$. This general fact implies that the motion of a particle under a central force takes place in a plane.

We can, therefore, use polar coordinates to describe the motion of the planet, with $\vec r = r \hat r$, and its velocity $\vec v = \dot r \hat r + r \dot \theta \hat \theta$. Therefore, the (conserved)
mechanical energy $E$ is given by
\begin{equation}
E = \frac{1}{2}mv^2 + V(r) = \frac{1}{2}m(\dot r^2 + r^2 \dot \theta^2) - \frac{GMm}{r}.
\end{equation}
The angular momentum is given by $\vec L = \vec r \times \vec p = mr^2 \dot \theta \hat z = L \hat z$, where $L = m r^2 \dot \theta$ is a constant. Using that $\dot \theta = L/mr^2$,
the energy can be written as
\begin{equation}
 E = \frac{1}{2}m\dot r^2 + \frac{L^2}{2 m r^2} - \frac{GMm}{r}.
\end{equation}
We complete the squares, obtaining
\begin{equation}
 E = \frac{m}{2}\dot r^2 + \frac{L^2}{2 m}\left(\frac{1}{r} - \frac{GMm^2}{L^2}\right)^2 - \frac{G^2M^2m^3}{2 L^2} .
\end{equation}
Introducing $K \equiv E + G^2 M^2m^3/2L^2$, we have
\begin{equation}
 K = \left( \sqrt{\frac{m}{2}} \dot r \right)^2 + \left( \frac{L}{\sqrt{2m}}\left(\frac{1}{r} - \frac{GMm^2}{L^2}\right)\right)^2 . \label{eqK}
\end{equation}
Since $K$ is the sum of two squares, it is a positive constant, unless $\dot r = 0$ and 
$1/r = GMm^2/L^2$, which corresponds to circular orbits. Therefore we can rewrite (\ref{eqK}) as
\begin{equation}
 \left(\sqrt{\frac{m}{2K}} \dot r \right)^2 + \left( \frac{L}{\sqrt{2mK}}\left(\frac{1}{r} - \frac{GMm^2}{L^2}\right)\right)^2 = 1 . \label{sumsquares}
\end{equation}
Equation (\ref{sumsquares}) allows us to write each of the squared terms as
\begin{eqnarray}
 \sqrt{\frac{m}{2K}} \dot r &=& \sin\phi , \label{sin} \\ \frac{L}{\sqrt{2mK}}\left(\frac{1}{r} - \frac{GMm^2}{L^2} \right) &=& \cos \phi \label{cos}
\end{eqnarray}
where $\phi$ is a function of time, to be determined. Taking the time derivative of the second expression in equation (\ref{cos}) and equation (\ref{sin}), we obtain
\begin{equation}
 \frac{\dot r}{r^2} =  \frac{\sqrt{2mK}}{L} \dot \phi \sin \phi = \frac{m}{L} \dot \phi \dot r , \label{LandPhi}
\end{equation}
implying $L = mr^2 \dot \phi$. But since $L = mr^2 \dot \theta$, we conclude that $\phi = \theta + \phi_ 0$. Without loss of generality we can take $\phi_0 = 0$, and solve for
$r$ in equation (\ref{cos}),
\begin{equation}
 r = \frac{p}{1 + e \cos\theta} , \label{keplerorbit}
\end{equation}
where $p = L^2/GMm^2$ and $e = \sqrt{2L^2K/G^2M^2m^3}$ is the eccentricity of the orbit. For completeness, we show that this is the equation of conic sections in polar coordinates.

We can rewrite equation (\ref{keplerorbit}) as
\begin{equation}
    r + e r\cos\theta = p , \label{polar}
\end{equation}
which implies
\begin{equation}
\sqrt{x^2 + y^2} + e x = p .
\end{equation}
From this we readily derive that
\begin{equation}
    x^2 \left(\frac{1-e^2}{e^2}\right) + \frac{2p}{e}x + \frac{y^2}{e^2} = \frac{p^2}{e^2} , 
\end{equation}
that is, for $e < 1$ it is an ellipse, for $e = 1$ a parabola and for $e > 1$ a hyperbole.

\section{The Kepler Problem as a Harmonic Oscillator}

We can rewrite Equation (\ref{eqK}) in such a way that we see a direct connection with the one dimensional harmonic oscillator. Using $L = m r^2 \dot \theta$ is a constant, we have
\begin{eqnarray}
  K &=& \frac{L^2}{L^2}\left( \sqrt{\frac{m}{2}} \dot r \right)^2 + \left( \frac{L}{\sqrt{2m}}\left(\frac{1}{r} - \frac{GMm^2}{L^2}\right)\right)^2 \nonumber \\
  \phantom{K} &=& \left( L\sqrt{\frac{m}{2}} \frac{1}{m r^2 \dot \theta}\dot r \right)^2 + \left( \frac{L}{\sqrt{2m}}\left(\frac{1}{r} - \frac{GMm^2}{L^2}\right)\right)^2 \nonumber \\
  \phantom{K} &=& \left({\frac{L}{\sqrt{2m}}} \frac{1}{r^2} \frac{dr}{d\theta} \right)^2 + \left( \frac{L}{\sqrt{2m}}\left(\frac{1}{r} - \frac{GMm^2}{L^2}\right)\right)^2 .
\end{eqnarray}
Finally, introducing $u = 1/r$, we have
\begin{equation}
 \left(\frac{du}{d\theta} \right)^2 + \left(u - \frac{GMm^2}{L^2}\right)^2 = \frac{2mK}{L^2},
\end{equation}
which is quadratic in $u$ and $\dot u$, being, formally, the same energy function as the one-dimensional harmonic oscillator. 

\section{A Related Problem}

Careful observations of the orbit of Mercury revealed it not to be an ellipse, as expected from a simple model of spherical celestial bodies \cite{Wells}. During it's orbit, Mercury's perihelion, the point nearest to the sun, precesses by a small amount. This led the french mathematician Urbain Le Verrier to conjecture the existence of a small planet, Vulcan, as it was called, with an orbit between Mercury and the sun. A first correction to the gravitational potential is proportional to $1/r^2$. We can use the same method to solve this problem, namely, consider the potential
\begin{equation}
V(r) = -\frac{GMm}{r} + \frac{\alpha}{r^2}
\end{equation}
where $\alpha$ is a constant. In this case, the mechanical energy is given by
\begin{equation}
    E = \frac{1}{2}m \dot r^2 + \frac{L^2}{2mr^2} - \frac{GMm}{r} + \frac{\alpha}{r^2} = \frac{1}{2}m \dot r^2 + \frac{\tilde L^2}{2mr^2} - \frac{GMm}{r}
\end{equation}
where $\tilde L^2 = L^2 + 2m\alpha$. Following the calculations in section \ref{KeplerProblem}, in particular, equation (\ref{LandPhi}) we find $\tilde L = m r^2 \dot \phi$, and therefore
\begin{equation}
     L\left(1 + \frac{2m\alpha}{L^2}\right)^{1/2} = m r^2 \dot \phi
\end{equation}
from which we derive $\phi = \theta \sqrt{1 + 2m\alpha/L^2}$, that is, the angular variable $\theta$ is multiplied by a constant, which depends on the total angular momentum. The solution is, similarly to the Kepler problem,
\begin{equation}
 r = \frac{p}{1 + e \cos(\theta \sqrt{1 + 2m\alpha/L^2})}
\end{equation}
where $p = {\tilde L} ^2/GMm^2$ and $e = \sqrt{2{\tilde L} ^2K/G^2M^2m^3}$, with $K \equiv E + G^2 M^2m^3/2{\tilde L}^2$ is the excentricity of the orbit. Notice that the perihelion happens at $\theta_n \sqrt{1 + 2m\alpha/L^2} = 2 n \pi$, $n = 0, 1, \ldots$. The angular separation between two perihelions is $\Delta \theta_n = 2\pi/\sqrt{1 + 2m\alpha/L^2}$, which shows the perihelion precession explicitly.

\section{Orbit under linear force}

We can use the same technique to solve the problem of the orbit of a point mass $m$ under the influence of a linear force, that is, a quadratic potential. The energy is given by
\begin{equation}
    E = \frac{1}{2} m \dot r^2 + \frac{1}{2}m r^2 \dot \theta^2 + \frac{1}{2}k r^2 ,
\end{equation}
where $k > 0$. The case $k < 0$ can be dealt in the same way or even obtained by taking $k \to -k$ in the final result. Since this is a central potential, the angular momentum $ L = m r^2 \dot \theta$ is conserved, and we have
\begin{equation}
    E = \frac{1}{2} m \dot r^2 + \frac{L^2}{2m r^2} + \frac{1}{2}k r^2
\end{equation}
We can complete the squares, like in the previous examples,
\begin{eqnarray}
    E &=& \frac{1}{2} m \dot r^2 + \frac{L^2}{2m r^2} + \frac{1}{2}k r^2 \nonumber \\
    \phantom{E} &=& \frac{1}{2} m \dot r^2 + \left( \frac{L}{\sqrt{2m}}\frac{1}{r} \pm \sqrt{\frac{k}{2}}r\right)^2 \mp L \sqrt{\frac{k}{m}} .
\end{eqnarray}
Introducing $\omega = \sqrt{k/m}$ and $K_\pm = E \pm L\omega$, we have
\begin{equation}
    \left(\sqrt{\frac{m}{2K_\pm}}\dot r\right)^2 + \left(\frac{L}{\sqrt{2mK_\pm}}\frac{1}{r} \pm \sqrt{\frac{k}{2K_\pm}}r \right)^2 = 1 .
\end{equation}
We can, therefore, write the terms inside the parentheses as $\sin \phi_\pm$ and $\cos \phi_\pm$, where $\phi_\pm$ is a function yet to be determined,
\begin{eqnarray}
  \sqrt{\frac{m}{2K_\pm}} \dot r &=& \sin\phi_\pm ; \\
  \frac{L}{\sqrt{2mK_\pm}}\frac{1}{r} \pm \sqrt{\frac{k}{2K_\pm}}r &=& \cos \phi_\pm .
\end{eqnarray}
Taking the time derivative of the second equation and using the first equation, we obtain
\begin{eqnarray}
    -\frac{L}{\sqrt{2mK_\pm}}\frac{1}{r^2} \dot r \pm \sqrt{\frac{k}{2K_\pm}} \dot r = - \dot \phi_\pm \sin \phi_\pm =  - \dot \phi_\pm \sqrt{\frac{m}{2K_\pm}} \dot r ,
\end{eqnarray}
that is,
\begin{equation}
     -\frac{L}{mr^2} \pm \omega = - \dot \phi_\pm  .
\end{equation}
Using $L=mr^2\dot \theta$ and integrating, we obtain
\begin{equation}
\phi_\pm = \theta \mp \omega t \pm \alpha_\pm
\end{equation}
where $\alpha_\pm$ are integration constants. We can choose $\theta=0$ at $t=0$. Moreover, we can also choose $\theta = 0$ as the point in the orbit when $\dot r =0$ (closest or farthest from the center of force), which corresponds to $\phi_\pm =0$. All this together implies we can take $\alpha_\pm = 0$ with the appropriate choice of coordinates and initial moment.

After some slightly tedious but straightforward algebra, we find
\begin{equation}
    r^2_\pm = \frac{2L^2\omega^2 +(E^2 - L^2 \omega^2) ( \sin^2 2 \theta \pm \sin^2 2\theta)}{2Ek+ 2k\sqrt{E^2-L^2\omega^2} \cos 2 \theta} .
\end{equation}
Let us analyze the solutions, starting with $r_+$,
\begin{equation}
      r^2_+= \frac{L^2\omega^2 +(E^2 - L^2 \omega^2) \sin^2 2\theta}{k(E+ \sqrt{E^2-L^2\omega^2} \cos 2 \theta)} = \frac{1}{k}(E - \sqrt{E^2 - L^2 \omega^2} \cos 2\theta) . \label{spurious}
\end{equation}
Plugging Equation (\ref{spurious}) in the energy expression for the linear orbit does not give a constant, therefore being a spurious solution which must be discarded.

A simpler way to convince oneself that Equation (\ref{spurious}) is not a solution of our problem is by noticing a peculiar behaviour, namely, for $E > 3 L\omega$, the curvature at $\theta = 0$, say, is negative, contradicting the fact that this is a central force: the motion can not be `curved' towards outside the trajectory. One qualitative way to see that is by inspecting the formal case $L = 0$. Here the polar equation describes two circles tangent at the origin, in an `8-shaped' figure. Increasing $L$ it is natural to  expect the negative curvature effect.

This leaves us with the `minus' solution for the physical motion of the particle
\begin{equation}
    r(\theta) = \frac{L\omega}{\sqrt{Ek + \sqrt{E^2-L^2\omega^2} k \cos 2 \theta}} .
\end{equation}

\section{Conclusions}

We obtained the complete solutions of the $1d$ harmonic oscillator and the central potential problems that depend on the radial coordinate as $1/r$, $1/r^2$ and $r^2$, using minimal technical background. The parameters of the solutions expressed directly in terms of the physical quantities, such as energy and angular momentum. We hope students, specially beginners, find this an appealing approach to such beautiful problems.

\begin{acknowledgments}
The author would like to thank Reinaldo de Melo e Souza and two anonymous referees, for helping improve this paper considerably.
\end{acknowledgments}

\end{document}